# Fluctuating diffusion-limited aggregates

Carlos I. Mendoza and Carlos M. Marques

*L.D.F.C.- UMR 7506, 3 rue de l'Université, 67084 Strasbourg Cedex, France*

(May 20, 2003)

## Abstract

We study the structure and growth of a diffusion-limited aggregate (DLA) for which the constitutive units remain mobile during the aggregation process. Contrary to DLA where far from equilibrium conditions are the prevalent factor for growth, the structure of the aggregate is here determined by a combination of annealed and quenched processes. The internal flexibility allows the aggregate to span the equilibrium configurational space, and such thermally driven motion further modifies the connectivity statistics of the growing branched structure.
61.43.Hv,05.40.-a,82.70.Dd

# I. INTRODUCTION

Diffusion-limited aggregation (DLA) [1] and other related models [2,3] have become paradigms for growth phenomena in far from equilibrium conditions [4]. In colloidal suspensions and other systems well described by these models [5], the elementary units that successively stick to form the aggregate remain thereafter at relative fixed positions, thus conferring an intrinsic rigidity to the structure: the statistical properties characterizing the disorder of the particle positions are quenched by the growth process itself. This is an important limitation that precludes application of diffusion-limited models to many aggregating systems, like the ones involving rearrangement within the clusters [6] or systems of polymer chains in solution [7], that consist of elements of a fluctuating nature. A realistic model for describing growth in this class of systems must include information not only about the colliding events that lead to the irreversible build up of the structure, but also about the equilibrium configurations that result from the brownian motion of the internally articulated constituents. Clearly, the statistical properties characterizing the positional disorder are here a combination of quenched and annelead processes but more crucially, the consecutive attachments of the incoming diffusive particles will now be biased by the fluctuations of the existing aggregate. In this letter we report numerical results on a model for diffusion-limited aggregation where, at the opposite of conventional DLA, the constituents remain freely articulated during and after the growing process.

# II. CONSTRUCTION OF A FLUCTUATING DLA

We build a fluctuating DLA (FDLA) aggregate from spherical monomers as sketched in Fig. 1. The growth process starts from a seed particle of size $r_0$ placed at the origin. Successive identical particles are attached to the aggregate in the following way: a monomer is released from a random point on a spherical shell that completely surrounds the aggregate. This monomer is allowed to move randomly while some of the monomers of the aggregate, chosen at random, are also allowed to move in random directions in steps of size $|\delta \mathbf{r}|$. If the distance between the released particle and any of the particles of the aggregate happens to be less than $r_0$ then the released particle sticks to that other particle at the contact point and a new one is released. If the particle moves to a distance too far away from the aggregate, this particle is "killed" and a new particle is released. In order to accelerate the simulation we used step size control. When the particle was at a distance $d$ to the cluster, a step of size $\delta = \max(d - 5r_0, 0.2r_0)$ was used. This value of the step size allows a rapid approach of the particle to the aggregate without hindering large excursions. Once attached to each other, the monomers interact through the conventional finitely extensible nonlinear elastic (FENE) potential [8], often used in polymer simulations:

$$V_b(r_{ij}) = -\frac{k}{2}(r_{sup} - r_0)^2 \ln\left[1 - \left(\frac{r_{ij} - r_0}{r_{sup} - r_0}\right)^2\right], \qquad (1)$$

where $k$ is the spring constant, and $r_{ij}$, $r_0$, and $r_{sup}$ are the instantaneous length of bond $ij$, the equilibrium bond length (or equivalently, the particle size), and the maximum bond length, respectively. The interaction between monomers not directly connected by a spring, is given by a hard sphere potential $V_{nb}(r_{ij}) = \infty$, if $r_{ij} \leq r_0$ ; $V_{nb}(r_{ij}) = 0$, if $r_{ij} > r_0$ where $r_{ij}$ is the instantaneous distance between particles $i$ and $j$. These potentials are used to accept or reject the movement of the internal monomers according to a standard Metropolis algorithm [9]. For the results presented in this letter each of the monomers of the aggregate moves on average once at each step of the incoming particle. In order to ensure that a full grown aggregate relaxes towards its equilibrium states, a non biased Montecarlo scheme is run at the end of the growth process. In all simulations, the system is held at fixed temperature $T = 1/k_B$, where $k_B$ is the Boltzman constant. The parameters used in the simulations for the FENE potential were: $k = 50$, $r_0 = 1$, and $r_{sup} = 1.2r_0$ to avoid bond crossing. The releasing sphere had size $r_{max} + r_{sup}$, where $r_{max}$ is the farthest distance between the origin and the monomers of the cluster, and the "killing" sphere had size $2r_{max}$. The size of the steps of the internal monomers was $|\delta \mathbf{r}|$ ($\delta r_\alpha \leq 0.2r_0$, $\alpha = x, y, z$). A typical snapshot of the aggregate is shown in Fig. 2. The number of MC steps used to relax the aggregate was of the order of $10^9$.

## III. RESULTS

We extracted pair correlation functions for the aggregates by the usual histogram method [10]. From a randomly chosen monomer, the distances $r$ to all other monomers are calculated and the histogram $N(r, \Delta)$ of the distance distribution is obtained by counting the number of distances in the interval $(r, r + \Delta)$. The histogram is further averaged for all the monomers of the aggregate and for a number of different runs. The pair correlation function was then obtained as $g(r) = N(r, r + \Delta)/(4\pi r^2 \Delta)$ and fitted to the stretched exponential form $g(r) = A r^{d_f - 3} \exp\{-(r/\xi)^a\}$, where $d_f$ is the fractal dimension and $\xi$ a decaying length associated with the size of the aggregate [11]. The exponent $a$ is larger than unity and $A$ is related to the total mass. A typical fit is shown in the inset of Fig. 3. In this figure, the variation of $d_f$ as a function of the total mass $N$ of the aggregate is shown for $N = 1000, 2000$ and $3000$. The fractal dimension was obtained from pair-correlation functions thermodynamically averaged for five different aggregates. It represents therefore a mixture of quenched and annealed averages, related respectively to the different connectivities of the aggregates and to the statistical thermodynamic conformations brought about by the brownian motion. Although aggregates larger than $N = 3000$ can be built in a reasonable computation time, the growth process with internal mobility leads to an aggregate that must still further relax in the bath – compare open to solid squares in Fig. 3. It is such time consuming after-growth relaxation process that puts an upper bound to the sizes we considered in this study.

The fractal dimension of the aggregate grows with the total particle number, as shown in Fig. 3. This is a familiar feature of DLA growth, as also shown in the figure, where the fractal dimensions of regular DLA is shown for comparison [12]. The pair-correlation functions were obtained in this case by averaging only over the quenched, topological disorder associated with the different connectivities of five different aggregates. For these, a reasonable asymptotic value is reached for $N \geq 3000$. For the FDLA values, the data does not show yet at $N = 3000$ a tendency to saturate indicating a probable asymptotic value at $d_f \geq 2.53$. Although one should bear in mind the relatively large error bars ($\Delta d_f/d_f \sim 5\%$) present in the determination of the fractal dimensions, our data show that the effect of the allowed internal mobility is to generate aggregates with *higher fractal dimension* than the ones generated by conventional DLA. Further confirmation that compacter objects are formed by this growth method can be seen from the Fourier transform of the pair correlation function, the so-called structure factor defined by

$$S(\mathbf{q}) = \frac{1}{N} \Big\langle \sum_{n,m=1}^{N} \exp\{i\mathbf{q} \cdot (\mathbf{r}_n - \mathbf{r}_m)\} \Big\rangle, \qquad (2)$$

where $N$ is the number of scattering units in the aggregate, $\mathbf{r}_i$ is the position of the $i$-th scattering unit, $\mathbf{q}$ is the wavevector and the quenched and annealed averages are denoted by $\langle ... \rangle$. Fig. 4 shows the structure factors of objects with $N = 2000$. They exhibit the characteristic shape of DLA structure factors, also similar to the scattering functions of star-like polymers and dendrimers and other computer-generated aggregates [13]. An initial low wavevector plateau is followed by a hump that crosses at higher wavevectors into a linear region of slope close to $-1.8$. The extension of the plateau is a measure of the size of the object, more precisely of its radius of gyration [14], $S(q \to 0) \sim 1 - q^2 R_G^2/3$ with $R_G^2 = 1/N \langle \sum_{n,m=1}^{N} (\mathbf{r}_n - \mathbf{r}_m)^2 \rangle$. From this measure the fractal dimension can also be extracted $N \sim R_G^{d_f}$. Graphically this implies that the fractal dimension can be obtained by the slope of the envelope of the humps of a series of structure factor curves of different masses. It is clear from Fig. 4 that the radius of gyration of a FDLA structure is smaller than the corresponding DLA quantity. The total mass of the FDLA structure is thus distributed over a smaller region, consistent with a denser object of higher fractal dimension. The local structure of the monomer correlations determines the high wavevector region of the scattering curves. Here, a self similar structure of dimension $\sim 1.8$ gives rise to the linear region of the curves shown in Fig. 4. Note also that this local structure is known to have a similar fractal dimension as that of diffusion-limited cluster aggregates (DLCA) [11].

## IV. DISCUSSION

Two effects might be at the origin of the increased compactness of the FDLA structure. One, related to its internal flexibility, can be better understood by considering first a directed growth process. Without internal mobility, that would lead to a rod-like object, with fractal dimension $d_f = 1$. Bringing mobility to this object would obviously transform it into a self-avoiding walk (SAW), with fractal dimension $d_f \simeq 1.7$, much larger than the original one. The SAW fractal dimension is the result of a compromise between excluded volume interactions that resist compaction, and conformational entropy that resists stretching [15]. If a linear growth process would prepare a quenched distribution of chains statistically stretched with respect to the SAW equilibrium distribution, the further introduction of internal degrees of freedom would anneal the distribution, leading to an increase of the fractal dimension. In order to see if this mechanism is at work for DLA structures, we first prepared conventional DLA aggregates of $N = 1000$, $2000$ and $3000$, then allowed them to thermodynamically relax under $10^9$ Monte-Carlo steps. Comparing the open and solid circles in Fig. 3 one sees that relaxation leads systematically to an increase of the fractal dimension. This means that the conventional DLA growth process does prepare quenched but statistically stretched structures. Further confirmation of this effect is provided in Fig. 5 where the average distance $r$ of a given monomer to the position of the seed monomer is shown as a function of the chemical distance $n$ between them. The chemical distance $n$ is defined as the number of elements connecting these two points. The slope of this curve is related to chain stretching along that chemical path [16]. As the figure shows, thermal relaxation is followed by a levelling of the curve close to the extremities of that particular branch, whereas the internal monomers retain their original stretching. This is a known steric effect also present in polymer brushes [17] or star polymers [18]. Close to the center, where the monomer density is higher, excluded volume interactions stretch branches outwards from their central attachment point. On the contrary, there is no force acting on the free ends of the branches. Chain stretching decreases therefore from its maximum at the center to a vanishing value at the free-end. It appears however that such mechanism is not sufficient to explain alone the observed compaction under FDLA growth, because, as shown in Fig. 3, the thermal relaxation of the conventional DLA stops short of the higher values obtained for FDLA.

A second effect that is likely to be at work compacting the final FDLA object, is related to its branch structure. Indeed, an incoming particle has a higher probability to stick first to an outer branch of the aggregate, than to penetrate deep, close to the seed monomer [1]. An useful indicator of the branch structure is provided by the average number of monomers at a chemical distance $n$ from the seed monomer. A monodisperse star, for instance, would have an uniform distribution with an upper cut-off corresponding to the arm length. A conventional DLA aggregate has a bell distribution. For the DLA aggregate of 2000 particles shown in Fig. 6 the maximum of the distribution corresponds to monomers separated from the seed by approximately 28 other monomers. The FDLA distribution shows that branches are shorter on average, the maximum being at approximately 20 monomers. The higher fractal dimension of FDLA is thus not only due to its internal flexibility but also to a different branching structure of the aggregate. Interestingly, the amount of free ends is quite constant during the growth process of both DLA and FDLA aggregates, at about 31%. This indicates that out of three new incoming monomers one contributes to a new branch being created, and the other two stick to existing free ends. This in turn implies that the more compact branch structure of FDLA is due to a deeper penetration of the incoming monomers.

## V. CONCLUSIONS

As a summary, we have shown that diffusion limited aggregation with soft particles that allow to internal mobility of the growing aggregate leads to more compact structures of higher fractal dimension. The increase compactness of the structure is due both to the internal relaxation of the growth structure but also to an increased compactness of its connectivity tree. We hope to extend in the near future our studies to larger aggregates: it remains yet to be shown if a well defined fractal dimension can be asymptotically reached.

## ACKNOWLEDGMENTS

This work was supported by the Chemistry Department of the CNRS, under AIP "Soutien aux Jeunes Equipes". It was also supported by a post-doctoral fellowship from the French MENRT.

# REFERENCES

[1] T.A. Witten, and L.M. Sander, Phys. Rev. Lett., 47 (1981) 1400.

[2] P. Meakin, Phys. Rev. Lett., 51 (1983) 1119.

[3] M. Kolb, R. Botet, and R. Jullien, Phys. Rev. Lett., 51 (1983) 1123.

[4] P. Meakin, *Fractals, scaling and growth far from equilibrium*, (Cambridge University Press, Cambridge) 1998.

[5] D.A. Weitz and M. Oliveria, Phys. Rev. Lett., 52 (1984) 1433.

[6] D.W. Schaefer, J.E. Martin, P. Wiltzius, and D.S. Cannell, Phys. Rev. Lett., 52 (1984) 2371.

[7] See for example, M.R. Gittings, Luca Cipelletti, V. Trappe, D.A. Weitz, M. In, and C. Marques, J. Phys. Chem. B, 104 (2000) 4381.

[8] K. Binder, A. Milchev, and J. Baschnagel, Ann. Rev. Mater. Sci., 26 (1996) 107.

[9] D. Frenkel and B. Smit, *Understanding molecular simulation*, (Academic Press, San Diego, California) 1996.

[10] M. Lach-hab, A.E. González, and E. Blaisten-Barojas, Phys. Rev. E, 57 (1998) 4520.

[11] C. Oh, and C.M. Sorensen, Phys. Rev. E, 57 (1998) 784.

[12] The value 2.45 for regular DLA is in reasonable agreement with the current value of 2.49. See L.M. Sander, Cont. Phys., 41 (2000) 203.

[13] R. Thouy and R. Jullien, J. Phys. I France, 6 (1996) 1365.

[14] J.S. Higgins and H.C. Benoît, *Polymers and neutron scattering*, (Oxford University Press, Oxford) 1996.

[15] P.G. de Gennes, *Scaling Concepts in Polymer Physics*, (Cornell University Press, Ithaca, New York) 1979.

[16] M. Doi and S.F. Edwards, *The Theory of Polymer Dynamics*, (Oxford University Press, Oxford) 1988.

[17] S.T. Milner, T.A. Witten, and M.E. Cates, Macromolecules, 21 (1988) 2610.

[18] M. Daoud and J.P. Cotton, J. Phys. France, 43 (1982) 531.

FIGURES

FIG. 1. Schematic sketch of a fluctuating DLA (FDLA) aggregate showing the constitutive units and the non-linear spring connecting them.

FIG. 2. Typical configuration of a FDLA aggregate of $N = 2000$ monomers.

FIG. 3. Fractal dimension $d_f$ as a function of the mass of the DLA aggregates. Circles refer to aggregates generated by the usual DLA algorithm, while the squares refer to the fluctuating diffusion-limited aggregation model (FDLA) explained in the text. Open symbols refer to non-relaxed structures while solid symbols correspond to structures having relaxed $10^9$ Monte-Carlo steps. The inset shows a typical pair-correlation function and the stretched exponential fit from which the fractal dimension is extracted.

FIG. 4. Scattering functions for aggregates built from DLA (dashed line) and FDLA (full line) growth processes. The dotted line corresponds to a conventional DLA structure relaxed after the growth process has taken place. All the aggregates have 2000 monomers.

FIG. 5. Distance of a given monomer to the seed monomer, as a function of the chemical distance between them: open circles correspond to DLA aggregates while solid circles refer to branches in FDLA aggregates. The slope of the curve is a measure of chain stretching. Results are for $N = 2000$.

FIG. 6. Distribution of the number of monomers $b(n)$ at a given chemical distance $n$ to the seed particle for aggregates of $N = 2000$. Dashed line corresponds to DLA whereas the solid line corresponds to FDLA. Here $< ... >$ means average over 5 different configurations.

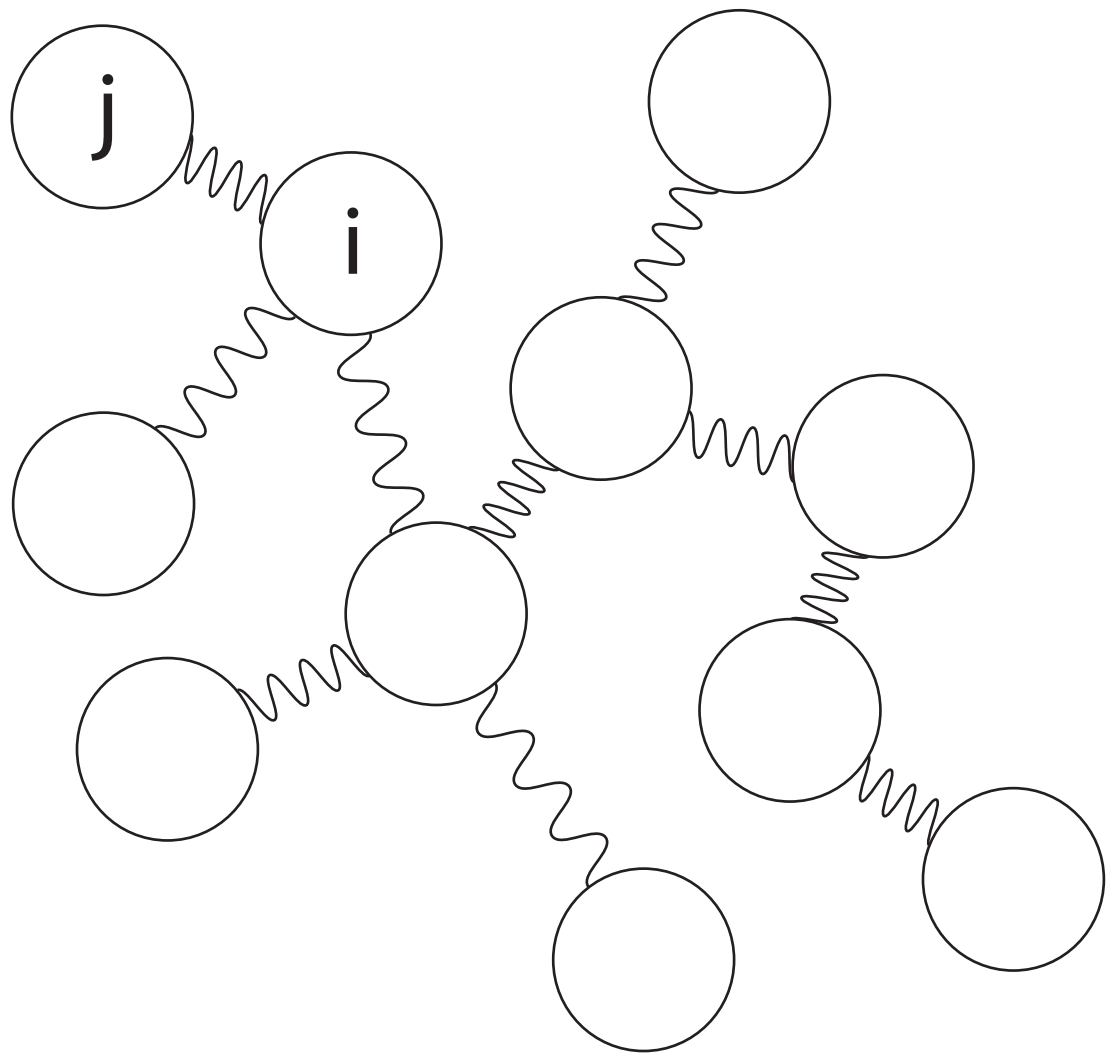

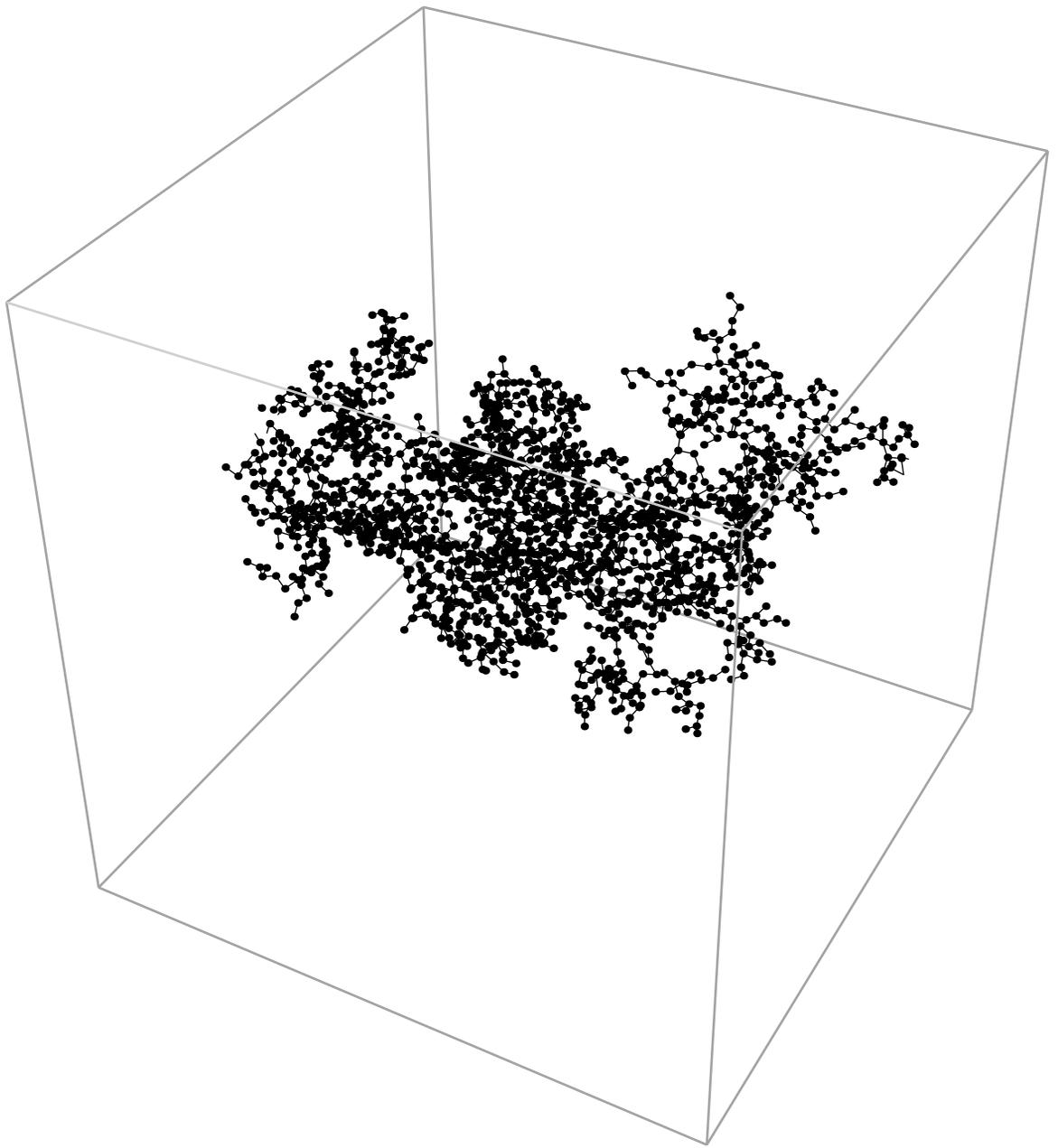

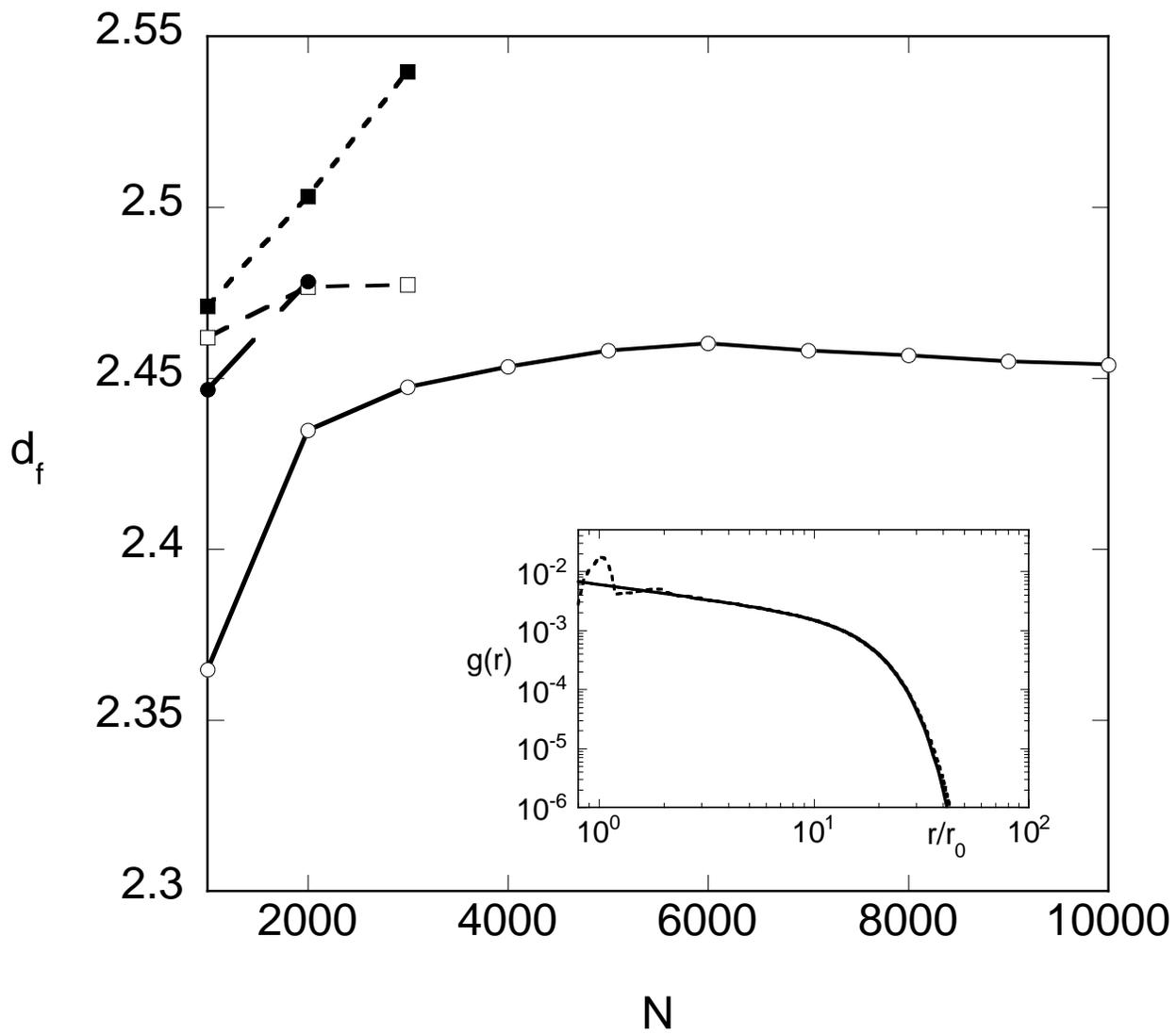

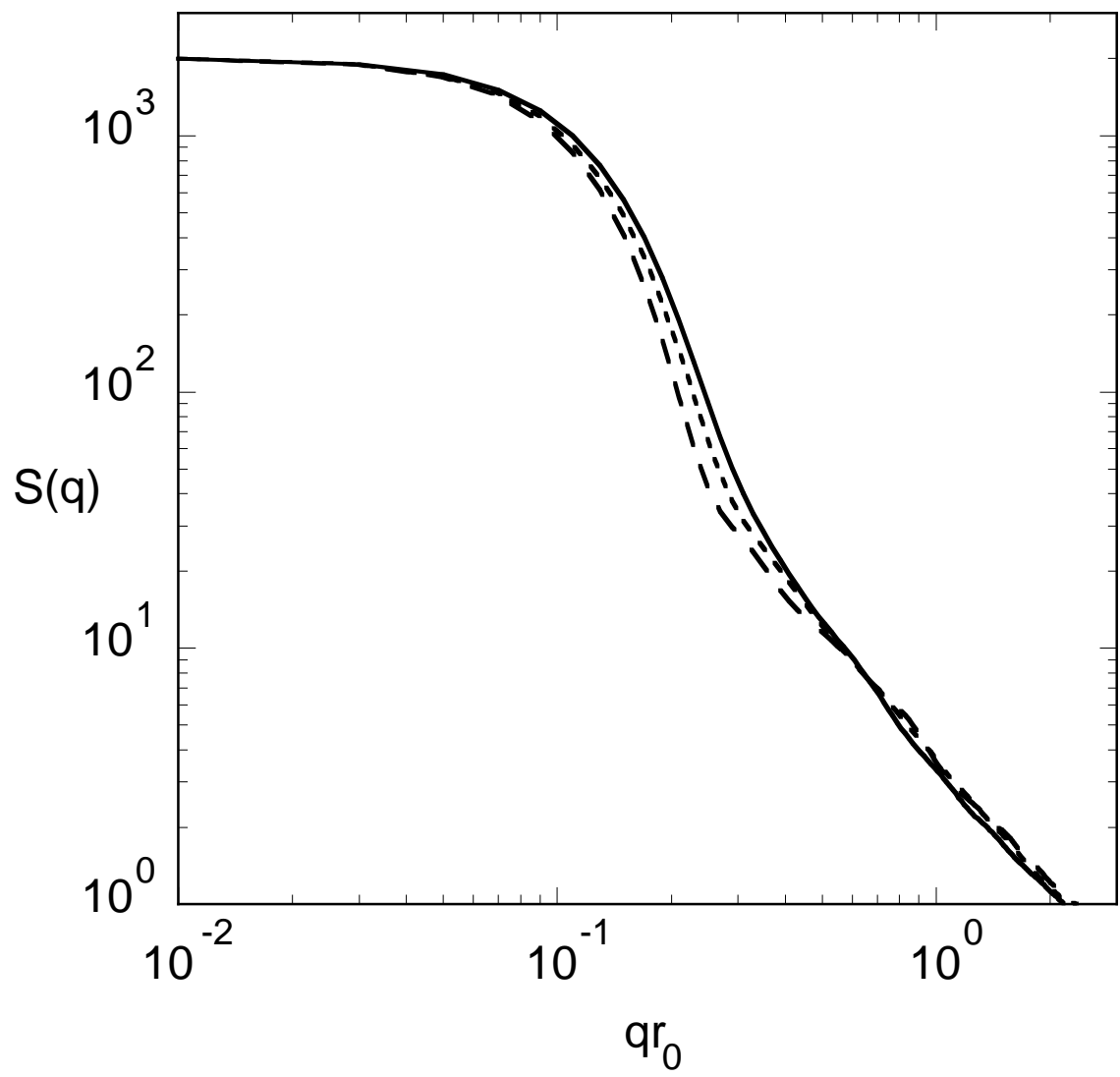

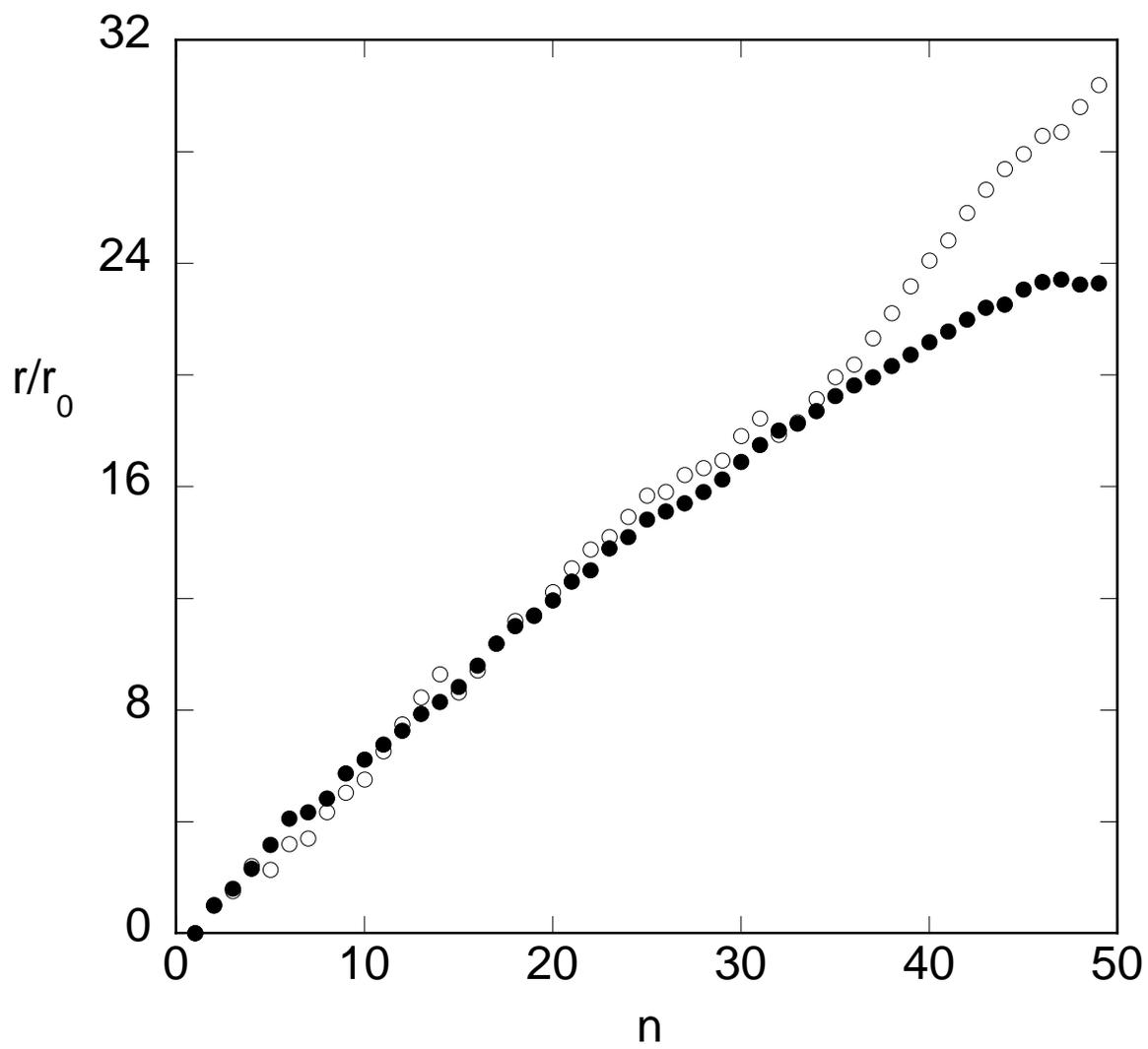

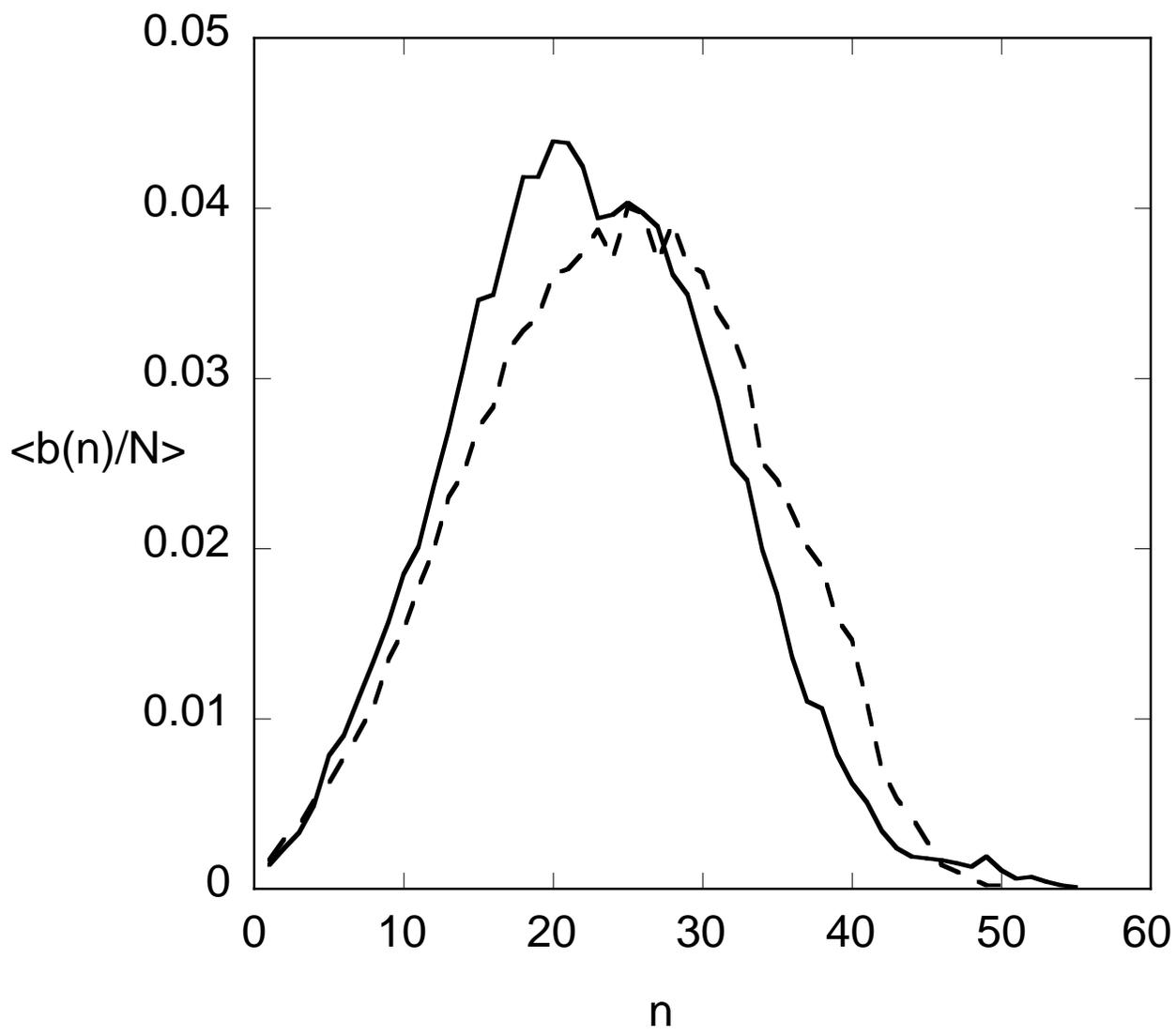